\documentclass{esannV2}
\usepackage{graphicx}
\usepackage[utf8]{inputenc}
\usepackage{amssymb,amsmath,array,url}
\usepackage{wrapfig}
\usepackage{subfig}
\usepackage{microtype}

\begin{document}
%style file for ESANN manuscripts
\title{Hierarchical clustering for graph visualization}

%***********************************************************************
% AUTHORS INFORMATION AREA
%***********************************************************************
\author{St\'ephan Cl\'emen\c{c}on$^1$, Hector De Arazoza$^2$,  Fabrice
  Rossi$^1$ and Viet-Chi Tran$^3$\footnote{This work was supported by
    the French Agency for Research under grant ANR Viroscopy
    (ANR-08-SYSC-016-03) and by AECID project D/030223/10}
%
% DO NOT MODIFY THE FOLLOWING '\vspace' ARGUMENT
\vspace{.3cm}\\
%
% Addresses and institutions (remove "1- " in case of a single institution)
1- Institut T\'el\'ecom, T\'el\'ecom ParisTech, LTCI - UMR CNRS 5141 \\
46, rue Barrault, 75013 Paris -- France
%
% Remove the next three lines in case of a single institution
\vspace{.1cm}\\
2- Facultad de Matem\'atica y Computaci\'on, Universidad de la Habana,\\
  La Habana, Cuba
\vspace{.1cm}\\
3- Laboratoire Paul Painlev\'e UMR CNRS No. 8524, Universit\'e Lille
  1,\\ 59 655 Villeneuve d'Ascq Cedex, France
}

\maketitle

\begin{abstract}
  This paper describes a graph visualization methodology based on hierarchical
  maximal modularity clustering, with interactive and significant coarsening
  and refining possibilities. An application of this method to HIV epidemic
  analysis in Cuba is outlined.
\end{abstract}

\section{Introduction}
Large graphs and networks are natural mathematical models of interacting
objects such as persons in social networks, articles in citation networks,
etc. \cite{Newman2003GraphSurveySIAM}. Unfortunately, the sizes of real
networks strongly limit visual exploratory analysis of their
properties. Indeed, despite the numerous graph visualization methods available
\cite{DiBattistaEtAl1999GraphDrawing,HermanEtAl2000Graph}, displaying a graph
with hundreds of vertices or more in a meaningful way remains difficult. It is
therefore common to rely on simplification techniques to draw a summary of a
complex graph: in general, a graph clustering algorithm is used to identify
clusters of nodes and the graph induced by the clustering is drawn
\cite{DBLP:conf/gd/EadesF96}. When the clustering is hierarchical, the full
graph can be represented using the structure imposed by the hierarchy: the
general layout is given by the higher level of the hierarchy in which the
graph is strongly simplified, then details are added in a top down way by
descending the hierarchy (see e.g., \cite{BOU07IV} for a recent algorithm of
this type).

Those approaches generally consider the (hierarchical) clustering of the graph
nodes to be obtained via one of the numerous graph clustering methods
available \cite{FortunatoSurveyGraphs2010,Schaeffer:COSREV2007}. In addition,
the significance of the hierarchical clustering is generally not questioned,
while a bad clustering could lead to serious interpretation errors. We propose
in this paper a methodology that addresses those two aspects. 

\section{Clustered graph construction}
This section described the proposed methodology which consists in building a
hierarchical graph clustering with guaranteed significance at each level,
using maximal modularity clustering as a building block.

\subsection{Maximal modularity clustering}
As show in recent surveys
\cite{FortunatoSurveyGraphs2010,Schaeffer:COSREV2007}, dozens of graph
clustering methods have been proposed. However, as pointed out in
e.g. \cite{RossiVilla2010TopoGraph}, not all of them are adapted to graph
visualization.  In seems indeed important to obtain clusters of nodes with a
high density of internal connections and a low number of external connections,
i.e., to find what is frequently called \emph{communities}. Among the
corresponding clustering methods, the ones that maximize the \emph{modularity}
of the obtained partition seem the more adapted: modularity maximization has
been shown to produce interesting partitions in many applications
\cite{FortunatoSurveyGraphs2010} and maximizing the modularity is equivalent
\cite{Noack2009}, to some extend, to optimizing the energy models used in many
graph visualization method, such as the standard Fruchterman Reingold
algorithm \cite{FruchtermanReingoldGraph1991}. 

While maximizing the modularity is a NP hard problem, high
quality sub-optimal solutions can be obtained efficiently, for instance using
deterministic annealing \cite{RossiVilla2010TopoGraph} or greedy methods such
as \cite{NoackRotta2009MultiLevelModularity}. Those latter methods have
reasonable computational complexities, proportional to the number of edges in
the graph multiply by the number of nodes: the implementation proposed in
\cite{NoackRotta2009MultiLevelModularity} can cluster a graph with $75\ 000$
nodes and $120\ 000$ edges in less than 2 minutes on standard PC hardware. 

\subsection{Clustering Significance}
While modularity is normalized and takes values between $-1$ and $1$, it is
\emph{a priori} difficult to tell whether a given value corresponds to an
actual significant partition of the graph or is just a manifestation of random
fluctuation. Indeed modularity maximization algorithms will generally discover
some partition in a random graph with no cluster structure. As clustered graph
visualization strongly relies on the hierarchical partitioning scheme, only
significant partitions should be considered. This could be done with the help
of theoretical results from
\cite{ReichardtBornholdt2007ModularityDistribution}. Indeed, the authors
compute the expected maximal modularity of random graphs with arbitrary degree
distribution. Given a graph, one can apply those results to its empirical
degree distribution to get a threshold value for the modularity; then if a
clustering has a larger modularity, it is assumed to be significant, i.e., to
represent a real structure in the graph.

This approach suffers from two drawbacks. Results from
\cite{ReichardtBornholdt2007ModularityDistribution} are only asymptotically
valid in the limit of large and dense graphs. In addition, only the expected
value is given, not the fluctuations around this value (e.g., the
variance). We propose therefore to rely on a simpler monte carlo strategy. We
built random graphs with no structure that have exactly the same degree
distribution as the one of the graph under study (this is the configuration
model random graph, see e.g., \cite{Newman2003GraphSurveySIAM}). Then we apply
a maximal modularity clustering algorithm (the one from
\cite{NoackRotta2009MultiLevelModularity}) to each of the random graph. The
obtained distribution of the modularity is valid under the null hypothesis of
no clustering and can be used to compute an estimate of the p-value of the
modularity of the clustering obtained on the original graph. This procedure is
time consuming, but current clustering methods are efficient enough to allow
it for medium scale graphs for e.g. 100 random trials. We recommend to
consider the clustering significant only if no random graph lead to a
modularity higher than the one of the original graph, i.e., for a p-value
lower than 1\%. For large scale graphs, we fall back to the approximation
provided in \cite{ReichardtBornholdt2007ModularityDistribution}.

\subsection{Hierarchical clustering}
To produce a clustered graph, we proceed as follows. We first compute a
maximal modularity clustering of the full graph and assess its significance
using the simulation strategy described above. If the clustering is not
relevant, then no meaningful representation can be done: this is reported to
the user as it means that the graph under study is better modelled as a random
network without structure than via clustering. 

Depending on the size of the graph and on the number of clusters produced by
the clustering algorithm, more coarsening and/or more refinement might be
useful. Following the general idea of clustered graph analysis, the clustering
is refined on a per cluster basis: we apply the same clustering method to each
of the clusters of the original graph, considered separately, that is
discarding inter-cluster connections. This can be repeated recursively if
needed, while keeping only significant clusterings. More precisely,
two significance levels are checked. We first apply the methodology described
above while considering the sub-graph in isolation. If the clustering is
significant, we also test its impact when integrated in the global clustering
of the graph: we make sure that, apart from the bottom level, each level of
the hierarchy contains a significant clustering of the original graph. 

Coarsening is done differently by applying the greedy strategy used by efficient
modularity maximization algorithms, in its simplest form: cluster pairs that
induce the least reduction in modularity are merged recursively as long as the
modularity of the obtained clustering remains above the significance level.

\section{Visualization}
The visualization of the clustered graph is based on 
classical solutions, such as \cite{BOU07IV}, with an important
difference. Indeed, efficient modularity maximization algorithms use generally
greedy coarsening techniques: any merge of the clusters of the best clustering
will reduce the modularity. Therefore, using the highest level of the
hierarchy to guide the display of the graph is likely to induce suboptimal
visualization (based on results of \cite{Noack2009}). Then we rely on the
following strategy.

A variant of the Fruchterman Reingold (FR) algorithm
\cite{FruchtermanReingoldGraph1991} is used to layout the quotient graph with
the best modularity: the repulsive force from a node is weighted by the size
of cluster it represents, rather than being uniform. This resulting display is
the original solution proposed to the user (see Figure
\ref{fig:clustered:graph} (a) for an example). Each node of this graph
represents a cluster, while edges are added between clusters whose members are
connected in the original graph. The surface of each node is proportional to
the cluster size. Colors (or gray levels) can be used to encode cluster
specific information, such as the internal connection density, or, as shown on
Figures \ref{fig:clustered:graph} and \ref{fig:coarsened:graph}, to display
quantities related to the distribution of some node property inside the
cluster.

Then an interactive exploration
is provided. The user can request the refining of the clustering. This is done
by descending in the hierarchical partition, down to the last level that
provide a significant clustering. At each refinement step, the layout of the
previous level is used for non refined nodes, while the FR algorithm is
applied to the refined node (see Figure \ref{fig:clustered:graph} (b)).

Coarsening is done in a similar but simpler way: the position of the node that
replaces nodes merged from the previous level is obtained as the center of mass
of those nodes, weighted by the size of the associated clusters (see Figure
\ref{fig:coarsened:graph}). 

\section{Application to the Cuban HIV-AIDS epidemic}
The proposed methodology is used to explore the largest connected component of
a graph of sexual contacts between patients infected by the HIV in Cuba (see
\cite{Auvert} for details). The corresponding database has been obtained via
the infection tracing methodology: when an individual is tested as HIV positive,
she/he is asked to name her/his sexual partners in the last two years. The
sexual contact is recorded for each partner already in the database. Others
are contacted and proposed a diagnosis. Positive new cases are added to the
database and the tracking proceeds. The database includes important
information about the patients, including there sexual orientation, age at
detection, etc. Up to 2006, the database contains almost 5 400 persons, among
which 2 386 belong to a single connected component. Applied to the
corresponding graph, classical visualization algorithms provide no insight on
the epidemic.

% \begin{wrapfigure}[16]{r}{0.5\linewidth}
% \vspace{-4ex}
%    \begin{center}
%      \includegraphics[width=0.99\linewidth]{clustered-chisq-so.pdf}
%    \end{center}
% \vspace{-4ex}
%    \caption{Initial layout}\label{fig:upper:level}
% \vspace{-1ex}
%  \end{wrapfigure}

\begin{figure}
  \centering
\subfloat[Best partition]{\includegraphics[width=0.45\linewidth]{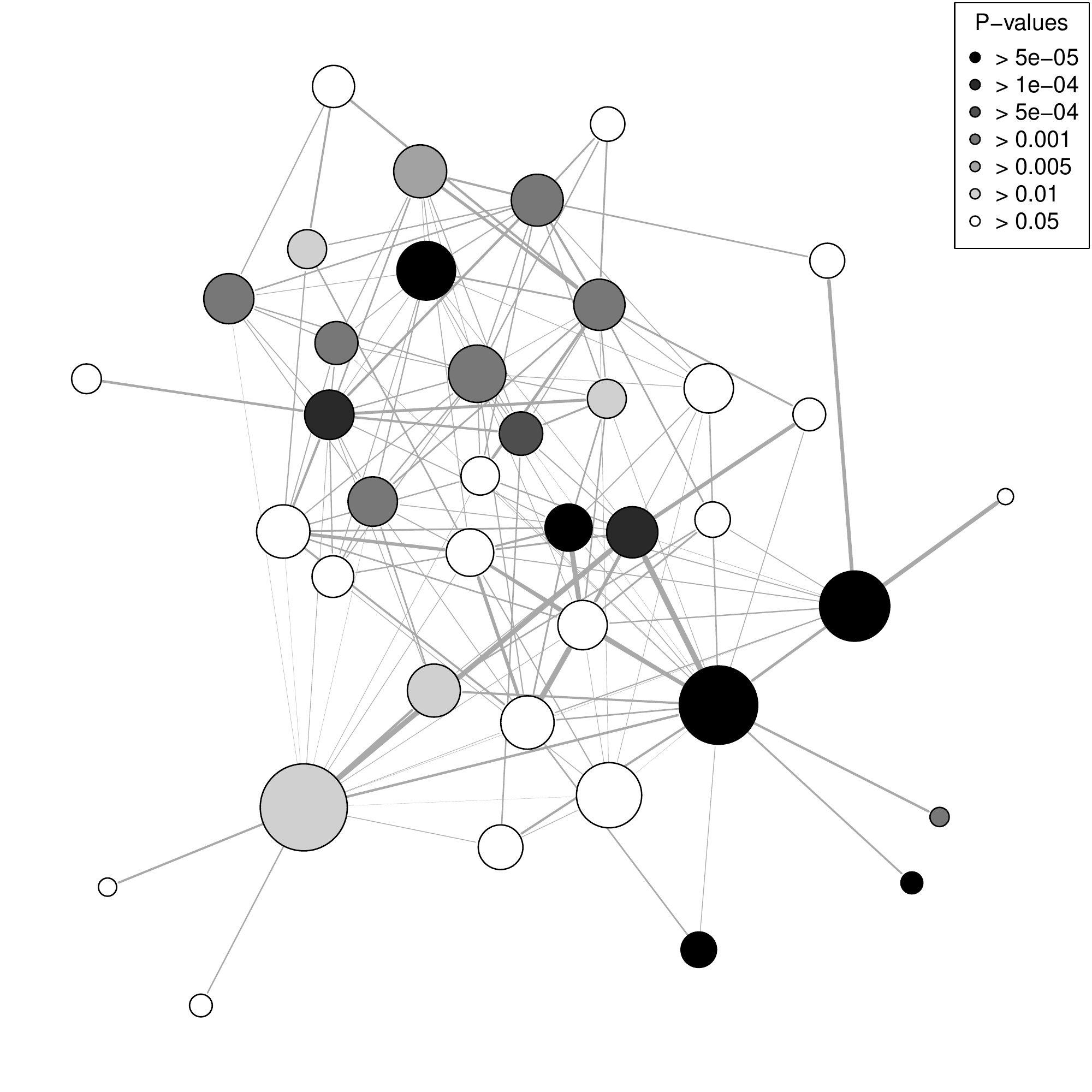}}%
\quad \subfloat[Maximally refined partition]{\includegraphics[width=0.45\linewidth]{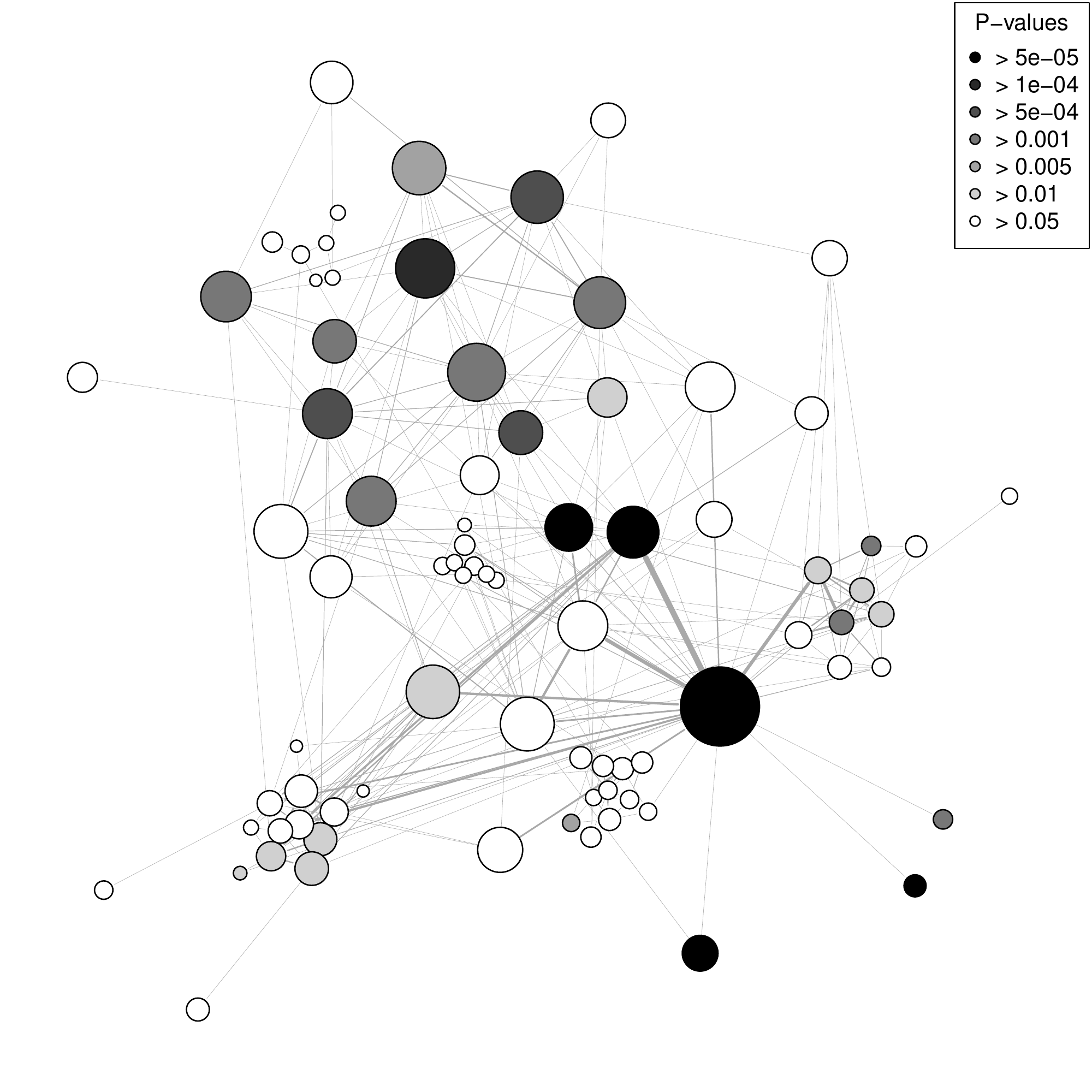}}  
\caption{Clustered graph visualization}
  \label{fig:clustered:graph}
\end{figure}

 Using \cite{NoackRotta2009MultiLevelModularity}, we obtain a partition of the
 graph into 39 clusters, with a modularity of 0.85. This is significantly
 higher than the modularities of random graphs with identical sizes and degree
 distributions: the highest value among 50 random graphs is 0.74. The
 corresponding layout is given by Figure \ref{fig:clustered:graph} (a): in this
 display nodes are darkened according to the p-value of a chi-squared test
 conducted on the distribution of the sexual orientation of persons in each
 cluster versus the distribution of the same variable in the full connected
 component. It appears clearly that some clusters have a specific distribution
 of the sexual orientation variable.

 The possibility of refining the clustering in this case is quite limited:
 only 5 of the 39 clusters have a significant substructure. Nevertheless,
 Figure \ref{fig:clustered:graph} (b), which shows the fully refined graph
 (with modularity 0.81) gives interesting insights on the underlying
 graph. For instance, an upper left gray cluster is split into 6 white
 clusters: while the best clustering of those persons lead to an atypical
 sexual orientation distribution, this is not the case of each
 sub-cluster. This directs the analyst to a detailed study of the
 corresponding persons: it turns out that the cluster consists mainly of
 bisexual male patients, who represent 76~\% of the original connected
 component. Sub-clusters are small enough ($\sim$ 7 patients) for bisexual
 male dominance to be possible by pure chance, while this is far less likely
 for the global cluster with 41 patients (among which 39 are bisexual males).

\begin{figure}
  \centering
\subfloat[Chi square P values]{\includegraphics[width=0.45\linewidth]{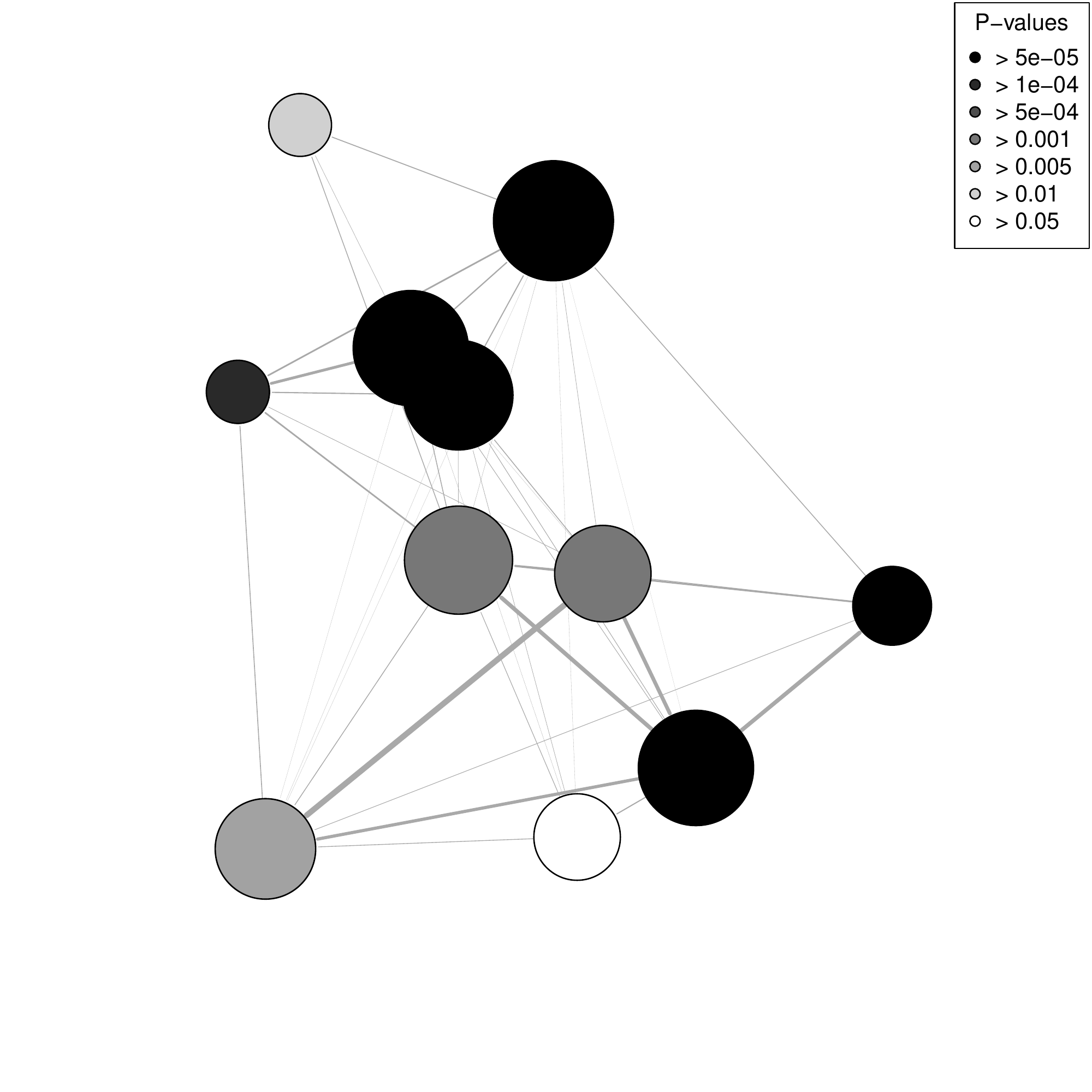}}%
\quad \subfloat[Pearson's residuals]{\includegraphics[width=0.45\linewidth]{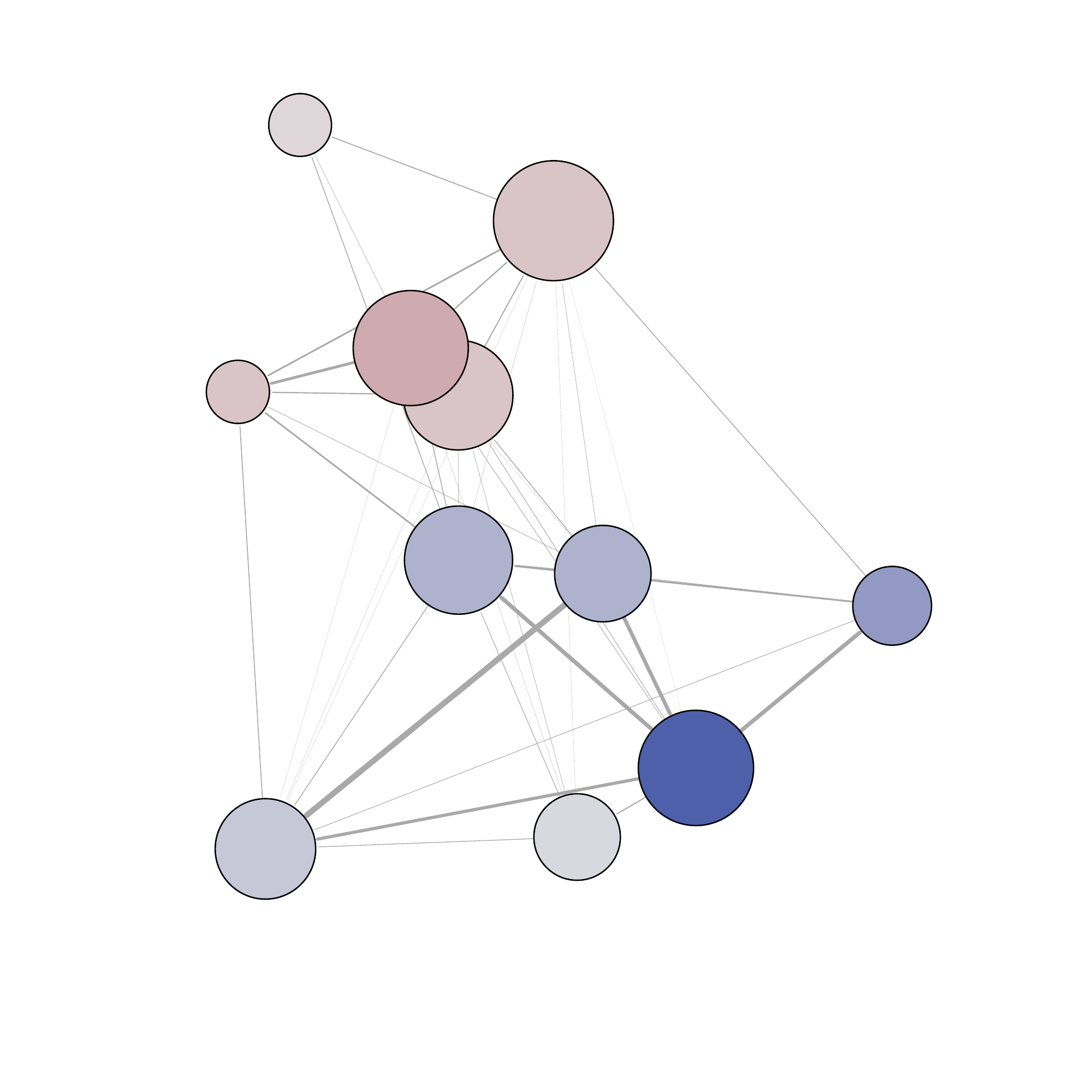}}  
\caption{Coarsened clustered graph visualization}
  \label{fig:coarsened:graph}
\end{figure}

Coarsening can be done more aggressively on this graph: clusterings down to 8
clusters have modularity above the random level. With 11 clusters, the
modularity reaches 0.81, a similar value as the maximally refined graph. While
Figure \ref{fig:clustered:graph} (a) is legible enough to allow direct
analysis, the coarsening emphasizes the separation of the graph into two
sparsely connected structures with mostly atypical sexual orientation
distributions in the associated clusters, as shown in Figure
\ref{fig:coarsened:graph} (a). Figure \ref{fig:coarsened:graph} (b) represents
the Pearson's residuals of the chi square tests for the ``bisexual male''
sexual orientation: it clearly shows that a part of the largest connected
component contains more than expected bisexual males (red nodes) while the
other part contains less than expected (blue nodes). Further analysis of those
data confirm the existence of a mostly homosexual/bisexual recent epidemic and
of an older and more heterosexual epidemic involving women.

\section{Conclusion}
The proposed methodology leverages progress in graph clustering, especially
the availability of fast modularity maximization algorithms. As shown on the
Cuban infection network, the visual results lead to important insight on the
graph content. Further work includes extensive comparisons to alternative
methods and combination of the proposed hierarchical clustering approach with
other clustered graph visualization approaches. 

\small
\bibliographystyle{abbrv}
\bibliography{graphvis}

\end{document}